\documentclass[aps,pra,twocolumn,groupedaddress]{revtex4-1}
\usepackage[dvipdfmx]{graphicx}
\usepackage{amsmath,amssymb}
\usepackage{bm}
\usepackage{braket}

\newcommand{\Arctan}{\text{Arctan}}

\renewcommand{\Im}{\text{Im}}
\newcommand{\eff}{\text{eff}}

\begin{document}


\title{Periodically-driven Kondo impurity in nonequilibrium steady states}
\author{Koudai Iwahori}
\email[]{iwahori@scphys.kyoto-u.ac.jp}
\author{Norio Kawakami}
\affiliation{Department of Physics, Kyoto University, Kyoto 606-8502, Japan}

\date{\today}

\begin{abstract}
We study the nonequilibrium dynamics of a periodically-driven anisotropic Kondo impurity model.
The periodic time dependence is introduced for a local magnetic field which couples to the impurity spin and also for an in-plane exchange interaction.
We obtain the exact results on the time evolution for arbitrary periodic time dependence at the special point in the parameter space known as the Toulouse limit. We first consider a specific case where the local magnetic field is periodically switched on and off.
When the driving period is much shorter than the inverse of the Kondo temperature, an intriguing oscillating behavior (resonance phenomenon) emerges in the time average of the impurity spin polarization with increasing the local magnetic field intensity.
By taking the high frequency limit of the external driving, we elucidate that the system recovers the translational invariance in time and can be described by a mixture of the zero-temperature and infinite-temperature properties.
In certain cases, the system is governed by either the zero-temperature or infinite-temperature properties, and therefore can be properly described by the corresponding equilibrium state.
\end{abstract}


\maketitle

\section{Introduction}
Single impurity in a fermionic or bosonic environment has been investigated intensively over the past several decades in solid state physics,
and many important concepts such as Anderson's orthogonality catastrophe, Kondo effect, polarons have been introduced.
As well as bulk solid-state systems, ultracold atomic systems and nanostructured systems such as quantum dots coupled to some leads have also provided intriguing platforms to study the single impurity problems thanks to recent progress in experimental techniques.
In nanostructured systems, the Kondo effect has been observed in transport properties \cite{Goldhaber-Gordon1998,Goldhaber-Gordon1998-2,Cronenwett1998,Schmid1998}. More recently, polaronic systems have been experimentally realized in ultracold atomic systems for a strongly imbalanced mixture of two species of atoms \cite{Schirotzek2009,Nascimbene2009,Will2011,Koschorreck2012,Kohstall2012}, 
and the realization of Kondo physics with ultracold atomic gases has been proposed theoretically \cite{Bauer2013,Nishida2013,Nakagawa2015}.
Because ultracold atomic systems and nanostructured systems have high controllability,  more detailed studies of equilibrium properties and also nonequilibrium properties of impurity systems, which are difficult to study in bulk solid-state systems, have been performed both experimentally \cite{Catani2012,Fukuhara2013,Fukuhara2013a,Hild2014,Kogan2004,Latta2011} and theoretically
\cite{Nordlander1999,Lobaskin2005,Anders2005,Anders2006,Hettler1995,Ng1996,Schiller1996,Goldin1997,Nordlander1998,Lopez1998,Goldin1999,Kaminski1999,Kaminski2000,Lopez2001,Heyl2010,Knap2012,Knap2013}.


In particular, real-time dynamics of the standard impurity models, such as the Anderson model and the Kondo model, has been studied intensively in nonequilibrium conditions; dynamics under periodic driving \cite{Hettler1995,Ng1996,Schiller1996,Goldin1997,Nordlander1998,Lopez1998,Goldin1999,Kaminski1999,Kaminski2000,Lopez2001,Heyl2010} as well as post-quench dynamics (i.e. dynamics after a sudden change of the system parameters) \cite{Nordlander1999,Lobaskin2005,Anders2005,Anders2006} have been explored. These studies have focused on nanostructured systems in which some leads and quantum dots are coupled with each other, and therefore have concentrated on the transport quantities such as the current between two leads through a dot. We will investigate below the time-dependent local quantities such as the impurity spin polarization and the spin density in a fermionic bath. This is partly motivated by recent experimental advances in cold atomic systems which enable us to observe the time-dependent local densities, for example, by using the single-site-resolved imaging technique in ultracold atomic systems \cite{Fukuhara2013,Fukuhara2013a,Hild2014}. Also, we note that these quantities have not been investigated in detail in the previous studies focused on the nanostructured systems. We will elucidate nontrivial time-dependent phenomena, especially a novel resonance phenomenon, emerging in the long-time limit of a periodically driven Kondo system.

More importantly, the detailed analysis of such real-time dynamics of the Kondo system provides an opportunity to address a fundamental problem of current interest in condensed matter physics, that is, how the periodically driven quantum systems behave in the long time limit. This problem has attracted much attention recently because novel properties of matter emerge by applying time-periodic external field to quantum systems \cite{Oka2009,Lindner2011,Kitagawa2011,Aidelsburger2013,Miyake2013}. Actually, recent studies have revealed intriguing properties of periodically-driven {\it isolated} systems where the Magnus expansion \cite{Magnus1954,Blanes2009} (a series expansion in the inverse of the external driving frequency) is not guaranteed to converge \cite{DAlessio2013,Lazarides2014b,Ponte2015,Mori2015,Abanin2015a,Kuwahara2016,Mori2016}. On the other hand, for {\it open} quantum systems, there are many issues to be explored. One of the problems is whether periodically-driven open systems can be described by some equilibrium systems \cite{Kohler1997,Breuer2000,Kohn2001,Hone2009,Ketzmerick2010,Iadecola2013,Langemeyer2014,Iadecola2015b,Shirai2015,Liu2015,Iadecola2015a,Shirai2016}. Recently, it has been shown that the Gibbs distribution of the Floquet states can emerge only when certain special conditions are fulfilled \cite{Shirai2015,Liu2015,Iadecola2015a,Shirai2016}. In general, an asymptotic state of the periodically driven open systems is not described by the Floquet-Gibbs state, so that it is important to study such open quantum systems in more detail to figure out in what conditions equilibrium-state properties emerge.

To address the above-mentioned problems, in this paper, we study the steady state of a periodically driven anisotropic Kondo impurity which is coupled to a fermionic bath with an infinite band width. The time dependence we consider is for the local magnetic field which couples to the impurity spin and also for the in-plane interaction strength.
We obtain the exact expression of time evolution for arbitrary time dependence of them by using the exact solution known as the Toulouse limit \cite{Toulouse1969}.
First, when the local magnetic field is periodically switched on and off and the in-plane interaction is time independent, 
we find a non-monotonic behavior of the time average of the impurity spin polarization, which is regarded as a resonance phenomenon, as a function of the intensity of the local magnetic field.
Second, by taking the high frequency limit of the external driving, we show that the system recovers time translational invariance and the properties of the system are described by a mixture of the zero-temperature state and the infinite-temperature state in general.
In certain cases, the system has the properties either at zero or infinite temperature, and thereby can be described by the equilibrium state in these special cases.

This paper is organized as follows.
In Sec. \ref{Sec_ModelMethod}, we summarize the Kondo model at the Toulouse limit and the method to obtain the exact analytical expression of the time evolution in the Heisenberg picture when the time periodic local magnetic field is applied to the system.
In Sec. \ref{Sec_RectangularCase}, we investigate a specific case where the local magnetic field is periodically switched on and off.
 In Sec. \ref{Sec_TimeDepInt}, we extend the calculations in Sec. \ref{Sec_ModelMethod} to the case where the in-plane interaction strength is also time dependent. This model enables us to address the question posed in the introduction: in what conditions equilibrium-state properties emerge for an open system. For this purpose, we consider the case where the local magnetic field and the in-plane interaction are both time dependent, and analyze the properties of the high frequency limit of the external driving. A brief summary of our results is presented in Sec. \ref{Sec_Conclusion}.

\section{Model and Method \label{Sec_ModelMethod}}
We consider a periodically-driven impurity spin coupled to a fermionic bath via the anisotropic Kondo exchange interaction,
\begin{equation}
	\begin{aligned}
		H(t) = \sum _{\sigma} & \int dx : \psi ^{\dagger} _{\sigma} (x) \big( -iu \partial _x \big) \psi _{\sigma} (x) : \\
							  & +\sum _{i=x,y,z} J_i S^i s^i (0) -h(t)S^z.
	\end{aligned}
\end{equation}
The operator $  \psi _{\sigma} (x) \ (\sigma = \uparrow , \downarrow) $  annihilates a fermion in the bath and the colons $: \cdots :$ denote the normal ordering.
$s^i (x) = \sum _{ss^{\prime}} :\psi ^{\dagger} _s (x) \sigma ^i _{ss^{\prime}} \psi _{s^{\prime}}(x): \ (i=x,y,z)$ are the spin density of the fermionic bath, where $\sigma ^i _{ss^{\prime}}$ are the Pauli matrices.
$ S^i \ (i=x,y,z)$ are the impurity spin operators whose spin is 1/2 and the couplings $J_i \ (i=x,y,z)$ have anisotropy: $J_x =J_y = J_{\perp},\ J_z=J_{\|}$.

We consider a time-dependent local magnetic field with the periodicity $h(t + \tau)=h(t)$ and assume that the initial state is in the ground state of the Kondo model without a local magnetic field.
Because the scattering term is point like and only fermions whose angular momentum is zero (s-wave) are scattered, the effective model is reduced to a one-dimensional system \cite{Affleck1991a}.
To simplify the problem, we take the strength of $z$-component coupling as $ J_{\|}/\pi u = \sqrt{2} (\sqrt{2}-1) $ (Toulouse limit \cite{Toulouse1969,Zarand2000}).
It is known that the exact results can be obtained at this point while keeping most of essential properties of the Kondo model intact.
At this point, we can transform the Kondo model to a non-interacting resonant level model by using the bosonization/refermionization methods \cite{Giamarchi2003}:
\begin{equation}
	\begin{aligned}
		H(t) = & \int dx : \tilde{\psi}^{\dagger} (x) \big( -iu \partial _x \big) \tilde{\psi} (x): \\
				& + \frac{J_{\perp}}{\sqrt{2 \pi \alpha}} \big( \tilde{\psi} ^{\dagger} (0) \tilde{c}_d + \text{h.c.} \big) -h(t) \big( \tilde{c}_d ^{\dagger} \tilde{c}_d -\frac{1}{2} \big) ,
	\end{aligned}\label{Eq_ResonantLevel}
\end{equation}
where the newly introduced operators $ \tilde{\psi} (x) $ and $ \tilde{c}_d $ are fermionic. 
The impurity spin polarization is represented by the occupation number of the resonant level, $S^z = \tilde{c}_d ^{\dagger} \tilde{c}_d -1/2 $.
The spin density of the fermionic bath is $s^z (x) = \sqrt{2}:\tilde{\psi}^{\dagger} (x) \tilde{\psi} (x): \ (x\neq 0)$.
The Kondo temperature, which provides a typical energy scale characterizing this quantum impurity system, can be determined by the impurity spin contribution to the specific heat as $T_K = \pi w J_{\perp} ^2/4 \pi \alpha u$, where $ w=0.4128 $ is known as the Wilson number \cite{Lobaskin2005}.

We calculate the time evolution of the system in the Heisenberg picture.
The time evolution operator of this system is $ U(t) = \mathcal{T} \exp \big[ -i \int _0 ^t du H(u) \big] $, where $ \mathcal{T} $ is the time-ordering operator. Because Hamiltonians of different time do not commute with each other, it is difficult to obtain the explicit expression for $ U(t) $.
Thus, we use the following strategy to obtain the time evolution of operators.
First, we divide the time-periodic local magnetic field $ h(t) $ into discrete $M$ time-steps:
\begin{equation}
	h(t) = \sum _{N=0} ^{\infty} \sum _{n=1} ^M \theta \big( t-(N-(n-1)/M) \tau \big) \theta \big( (N+n/M)\tau -t \big) h^{(n)}.
\end{equation}
When $ h(t) $ is divided into $M$ time-steps, the time evolution of operators in each period reads 
\begin{equation}
	\tilde{c}_l (\tau) = e^{i H^{(1)} \tau /M} \cdots e^{i H^{(M)} \tau /M} \tilde{c}_l e^{-i H^{(M)} \tau /M} \cdots e^{-i H^{(1)} \tau /M},
\end{equation}
where $ l=k,d $ and $ \tilde{c}_k $ is a Fourier coefficient of $ \tilde{\psi}(x) $ $( \tilde{\psi} (x) = 1/\sqrt{L} \sum _k e^{ikx} \tilde{c}_k )$.
$ H^{(n)} $ is a time-independent Kondo model with a local static magnetic field $ h^{(n)} $.
As the Hamiltonian $ H^{(n)} $ is quadratic, the time evolution by $ H^{(n)} $ is denoted as,
\begin{equation}
	\begin{aligned}
		c_l (t) &= e^{iH^{(n)}t}c_l e^{-iH^{(n)}t} \\
				&= \sum _{l^{\prime}} G^{(n)} _{ll^{\prime}} (t) c_{l^{\prime}} , \ (l=k,d) ,
	\end{aligned}
\end{equation}
where $ G^{(n)} _{ll^{\prime}} (t) $ is the transition matrix from $ l^{\prime} $ to $ l $.
Thus, the time evolution in a period is given by
\begin{equation}
	\begin{aligned}
		& \tilde{c}_l (\tau) = \sum _{l^{\prime}} M_{ll^{\prime}} \tilde{c}_{l^{\prime}} \\
		& M_{ll^{\prime}} = \sum _{ \{ l_i \} } G^{(M)} _{ll_1} \big(\tau /M \big) G^{(M-1)} _{l_1 l_2} \big(\tau /M \big) \cdots G^{(1)} _{l_{M-1} l^{\prime}} \big(\tau /M \big) .
	\end{aligned}
\end{equation}
The explicit expression of $ M_{ll^{\prime}} $ is shown in the Appendix \ref{appendix_a}.
By taking the continuum limit, the time evolution of the operators $ \tilde{\psi}(x) $ and $ \tilde{c}_d $ is obtained in the long-time limit,
\begin{equation}
	\begin{aligned}
		& \begin{aligned}
			\tilde{\psi} (x,t) = & \tilde{\psi}(x-ut) \\
								 & -i \sum _k \frac{J_{\perp}}{\sqrt{2 \pi \alpha u^2}} \theta (x) T_{dk} (t-x/u) \tilde{c} _k \\
		  \end{aligned} \\
		& \tilde{c}_{d}(t) = \sum _k T_{dk} (t) \tilde{c} _k \\
		& \begin{aligned}
			T_{dk} (t) & = \frac{J_{\perp}}{\sqrt{2 \pi \alpha L}} \frac{-ie^{-i\epsilon _k t}}{1-e^{i\epsilon _k \tau} M_{dd}} \\
						& \times \int _0 ^{\tau} ds \exp \Big[ i \int _{\tau -s} ^{\tau} h(s^{\prime} +t) ds^{\prime} - \Delta \tau +i \epsilon _k s \Big] \\
		  \end{aligned} \\
		& M_{dd} = \exp \Big[ i \int _0 ^{\tau} h(t) dt - \Delta \tau \Big] ,\ \Delta = \frac{J_{\perp} ^2}{4\pi \alpha u}.
	\end{aligned}
	\label{Eq_TransitionMatrix}
\end{equation}
These expressions are applicable for an arbitrary time-periodic function $h(t)$ and contains the whole information about the dynamics of the system including the fermionic bath. We note that similar exact analytical expressions for time-dependent current between the two leads through a nanostructure are obtained by using the Keldysh Green function method \cite{Wingreen1993,Schiller1996}.

The relation between the spin density of the fermionic bath and the time derivative of the impurity spin is also obtained:
\begin{equation}
	\braket{s^z(x,t)}=\begin{cases}
						-\frac{\sqrt{2}}{u}\braket{\frac{d}{dt} S^z (t-x/u)} & (x>0) \\[5pt]
						0													 & (x<0) 
					  \end{cases}
					  \label{Eq_SpinRelation}
\end{equation}
This relation means that a fermion whose spin is antiparallel to the impurity spin is scattered and propagates with the Fermi velocity $u$ 
because the fermionic bath and the impurity spin are coupled by antiferromagnetic interaction and the fermionic bath has a linear dispersion.

\section{Time-Dependent Properties and emergent Resonance Phenomenon  \label{Sec_RectangularCase}}
\subsection{Time evolution of observables \label{SubSec_RectangularTimeEvolution}}
\begin{figure*}
	\centering
	\includegraphics[width=150mm]{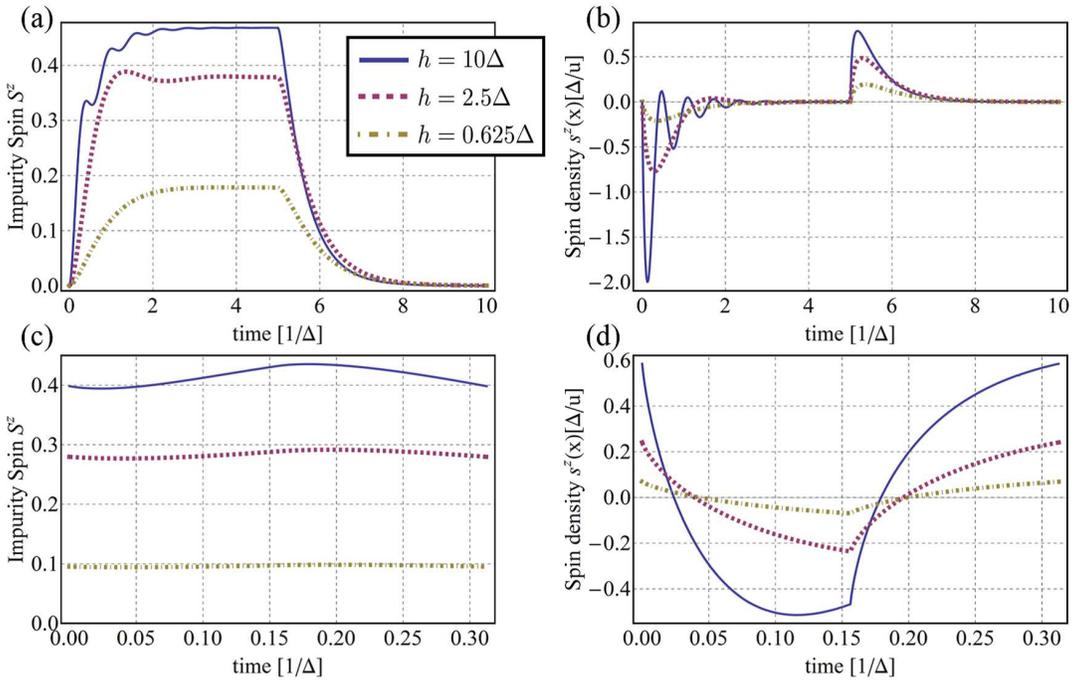}
	\caption{Time evolution of the impurity spin polarization (a), (c) and the spin density of the fermionic bath (b), (d) in a period for several choices of local magnetic field intensity $h$ 
			($h=10\Delta$, blue solid line; $h=2.5\Delta$, red dashed line; $h=0.625\Delta$, yellow dashed dotted line). 
			(a) and (b) are the figures for $ \tau = 10/\Delta $, while (c) and (d) are for $ \tau = 0.3125/\Delta $.\label{Fig_RectangularTimeEvolution}}
\end{figure*}
Here, we discuss a specific case: the local magnetic field is periodically switched on and off. Note that the following analysis and discussion can be straightforwardly applied to more generic periodically driven systems.
The time dependence of $ h(t) $ we consider is
\begin{equation}
	h(t) = \sum _{N=0} ^{\infty} \theta \big( t-N\tau \big) \theta \big( N\tau + \tau /2 -t \big) h .
\end{equation}
Decompose time into periodic intervals:	$t=N\tau +s , \ N \in \mathbb{N} , \ 0<s<\tau$.
Then, the time evolution of the impurity spin polarization in the steady state 
in the long-time limit 
is
\begin{equation}
	\braket{S^z (t)} \xrightarrow{N\rightarrow \infty} 
		\begin{cases}
			\begin{aligned}
				A_1 + A_2 & e^{-(T_K /\pi w)s} +\text{Re} \big[ A_3 (s) e^{ihs}] \\
						& (0<s<\tau /2)
			\end{aligned} \\
			\begin{aligned}
				A_1 ^{\prime} +A_2 ^{\prime} & e^{-(T_K /\pi w)(s-\tau /2)} \\
															& (\tau /2 <s<\tau).
			\end{aligned}
		\end{cases}
\end{equation}
The prefactors $ A_i ^{(\prime)} (s) $ can be obtained by the momentum integral in the fermionic-bath sector.
The second term is the relaxation term due to the switching of the field intensity.
The third term in $ 0<s<\tau /2 $ represents the oscillation caused by the local magnetic field, which comes from the interference between the impurity spin and the spins in the fermionic bath. Recall that
the spin density of the fermionic bath is represented by the time derivative of the impurity spin polarization (see Eq. (\ref{Eq_SpinRelation})).

The time evolution of the impurity spin polarization $ \braket{S^z (t)} $ and the spin density of the fermionic bath $ \braket{s^z(x,t)} $ in a period are shown in Fig. \ref{Fig_RectangularTimeEvolution}.
Because the impurity spin polarization $ \braket{S^z (t)} $ is time periodic in the steady state, the spin density of fermionic bath $ \braket{s^z(x,t)} $ is spatially periodic as well as time periodic, as seen from Eq. (\ref{Eq_SpinRelation}).
We thus show only the data at the point $x=Mu\tau \ (M\in \mathbb{N})$.
In the first half of the period the local magnetic field is switched on (i.e. $ h(t) = h $), while in the second half it is off (i.e. $ h(t) = 0 $).
In the long driving period (see the figures of $ \tau = 10 /\Delta $), the impurity spin polarization oscillates with a frequency proportional to the field intensity $h$, and eventually relaxes to its equilibrium value in the time scale proportional to $1/T_K$.  
Wave packets whose width is of the order of the inverse of the Kondo temperature are observed in the spin density of the fermionic bath since the spin density of the fermionic bath is given by the time derivative of the impurity spin polarization.
These wave packets propagate with the Fermi velocity $u$ without decaying due to the linear dispersion.
In the short driving period (see the figures of $ \tau = 0.3125 /\Delta$), the system cannot keep up with the temporal change of the local magnetic field and thus approaches constant values.
Note that a large cusp is observed at a half of the period in the spin density of the fermionic bath, because it is given by the time derivative of the impurity spin polarization and thus amplified by the factor $\Omega=2\pi /\tau$.


\subsection{Emergent resonance in the time averaged observables}
In the limit of fast driving, the impurity spin polarization $ \braket{S^z (t)} $ approaches a temporally constant value (see Fig. \ref{Fig_RectangularTimeEvolution}). Under this condition, we investigate how the time average of the impurity spin polarization, $ \overline{S^z} $, depends on the local magnetic field intensity $h$, and elucidate their intriguing resonance phenomenon. Here we define the time average $ \overline{S^z} $ and its variance $ \Delta S^z $:
\begin{equation}
	\begin{aligned}
		& \overline{S^z} = \lim _{N \rightarrow \infty} \int _0 ^{\tau} \frac{ds}{\tau} \braket{S^z (N\tau +s)} \\
		& \Delta S^z = \lim _{N \rightarrow \infty} \int _0 ^{\tau} \frac{ds}{\tau} \Big| \braket{S^z (N\tau +s)} -\overline{S^z} \Big| ,
	\end{aligned}
\end{equation}
where $ \lim _{N\rightarrow \infty} $ is taken to extract the steady state.

\begin{figure}[t]
	\centering
	\includegraphics[width=80mm]{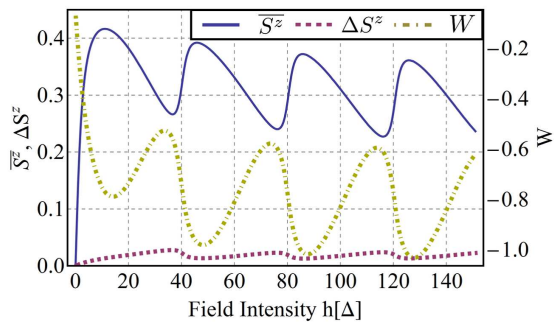}
	\caption{Time-averaged quantities as a function of the local magnetic field intensity $h$: Time-averaged impurity spin polarization $\overline{S^z}$ (blue solid line), Variance $\Delta S^z$ (red dashed line), Work $W$ (yellow dashed dotted line).\label{Fig_RectangularAverage}}
\end{figure}

The obtained results are shown in Fig. \ref{Fig_RectangularAverage} as a function of the field intensity $h$ for a choice of the driving period, $\tau=0.3125\Delta$. Naively, we expect that $ \overline{S^z} $ increases monotonically and saturates to $1/2$ with increasing $h$. Interestingly, however, an emergent oscillation is observed as a function of $h$ and its period is determined by the driving period of the local magnetic field. When $ h \gg T_K $, the time average of the impurity spin polarization $ \overline{S^z}(h) $ becomes periodic as a function of $h$
\begin{equation}
	\overline{S^z} (h+4\pi /\tau) = \overline{S^z}(h) \ (h \gg T_K) ,
\end{equation}
where $ \tau $ is the driving period of the local magnetic field.

To see the essence of this behavior, let us define the ``power" $P$ and the ``work done in a period" $ W $ for this system, based on an analogy to the power in classical mechanics,
\begin{equation}
	\begin{aligned}
		&P(t)=\frac{dS^z}{dt} \frac{dh(t)}{dt} \\
		&W= \lim _{N \rightarrow \infty} \int _0 ^{\tau} P(N \tau +s) ds .
	\end{aligned}
\end{equation}
It is expected that $ dS^z /dt $ corresponds to the velocity of a particle $v=dx(t)/dt$, and $ dh(t)/dt $ corresponds to the force $F$ in classical mechanics, that is, the following correspondence to classical mechanics is expected:
\begin{equation}
	\frac{dS^z (t)}{dt} \frac{dh(t)}{dt} \leftrightarrow \bm{v} \cdot \bm{F} = \frac{d}{dt} \text{(Kinetic energy)}
\end{equation}
The quantity $W$ is shown as a function of $h$  in Fig. \ref{Fig_RectangularAverage}. The emergent oscillation in the time average of the impurity spin polarization as a function of $h$ can be attributed to a resonance phenomenon between the oscillation of the impurity spin polarization (see Fig. \ref{Fig_RectangularTimeEvolution}) and switching of the local magnetic field.
The impurity spin polarization oscillates with the period of $ 2 \pi / h $ in the first half of the period, and the local magnetic field is switched off just at the half of the period.
Analogy to an oscillator in classical mechanics would suggest that if the time derivative of the impurity spin polarization is negative at the half of the period where the local magnetic field is switched off, the amplitude toward the negative direction would become larger, and therefore the time average would be smaller.
Note that the ``work done in a period" $W$ oscillates with the same period but an opposite phase, whereas the variance of the impurity spin polarization  $ \overline{S^z} $ is small but oscillates with the same phase as the ``work". These behaviors are indeed consistent with our interpretation. The resonance condition deduced by the analogy mentioned above is
\begin{equation}
	\frac{2 \pi n}{h} = \frac{\tau}{2} \Leftrightarrow h=\frac{4 \pi n}{\tau}  \ (n \in \mathbb{N}) ,
\end{equation}
which reproduces the period observed in Fig. \ref{Fig_RectangularAverage}.

\section{Some novel properties driven by time-dependent interactions\label{Sec_TimeDepInt}}
When the in-plane interaction $ J_{\perp} $ is also time-periodic, the system shows quite different behaviors from those without the in-plane interaction driving. The method in the Sec. \ref{Sec_ModelMethod} can be easily extended to the case where both the local magnetic field $h(t)$ and the in-plane interaction $J_{\perp}(t)$ are time-periodic functions.
The time evolution of the operators in this condition is shown in the Appendix \ref{appendix_a}, where the relation between the spin density of the fermionic bath and the time derivative of the impurity spin polarization (Eq. (\ref{Eq_SpinRelation})) also holds in this case.

In particular, we investigate below our driven system in the high frequency limit of the external driving and elucidate the characteristic long-time behavior of the periodically driven system.
Note that the Magnus expansion cannot be applied to this system due to the infinite band width and the Kondo model can be regarded as an open quantum system where a localized spin couples to a reservoir.
Therefore, it is instructive to study the properties of this system from the viewpoints of periodically driven open quantum systems as well as the divergence of the Magnus expansion. This is indeed related to the question whether the driven system can be described by some equilibrium states.

\subsection{Dynamics of the observables}
Here, we consider a specific case where the time dependence of the local magnetic field $h(t)$ and the in-plane interaction parameter $J_{\perp}(t)$ is given by
\begin{equation}
	h(t)=h\sin (2\Omega t) , \ J_{\perp}(t)=J_{\perp}\sin \Big( \Omega t +\frac{\pi}{4} \Big) .
\end{equation}
Note that the time dependence of the local magnetic field $h(t)$ equals that of the square of the in-plane interaction parameter $ J_{\perp} ^2 (t) $, which has the essence of dynamics when both the local magnetic field and the in-plane interaction parameter are time dependent.
The dynamics of the impurity spin polarization in the steady state is 
\begin{equation}
	\begin{aligned}
		& \braket{S^z(N\tau +s)} \xrightarrow{N \rightarrow \infty} e^{x \cos (2\Omega s)} A(s) -\frac{1}{2} \\
		& A(t) = \int _{-\infty} ^0 \frac{d \omega}{2 \pi} \Big| \sum _{p \in \mathbb{Z}} \frac{J_p (z)-J_{p+1} (z)}{\omega -(2p+1)\Omega /\overline{\Delta} +i} e^{2ip(\Omega t+\pi /4)} \Big| ^2 \\
		& \overline{\Delta} = \frac{1}{4\pi \alpha u} \int _0 ^{\tau} \frac{dt}{\tau} J_{\perp} ^2 (t) , \ x=\frac{\overline{\Delta}}{\Omega} , \ z=\frac{h+i\overline{\Delta}}{2\Omega} ,
	\end{aligned}
\end{equation}
where $J_n (z)$ is the integer Bessel function.
Because $A(t+\tau /2) = A(t)$, the impurity spin polarization is time periodic whose period is half of the driving period $\tau$: $ \braket{S^z(t+\tau /2)}=\braket{S^z(t)} $.
This is because the in-plane interaction parameter acts as an effective hopping parameter, and thereby the square of the in-plane interaction parameter appears in the observables.

\begin{figure*}[t]
	\centering
	\includegraphics[width=150mm]{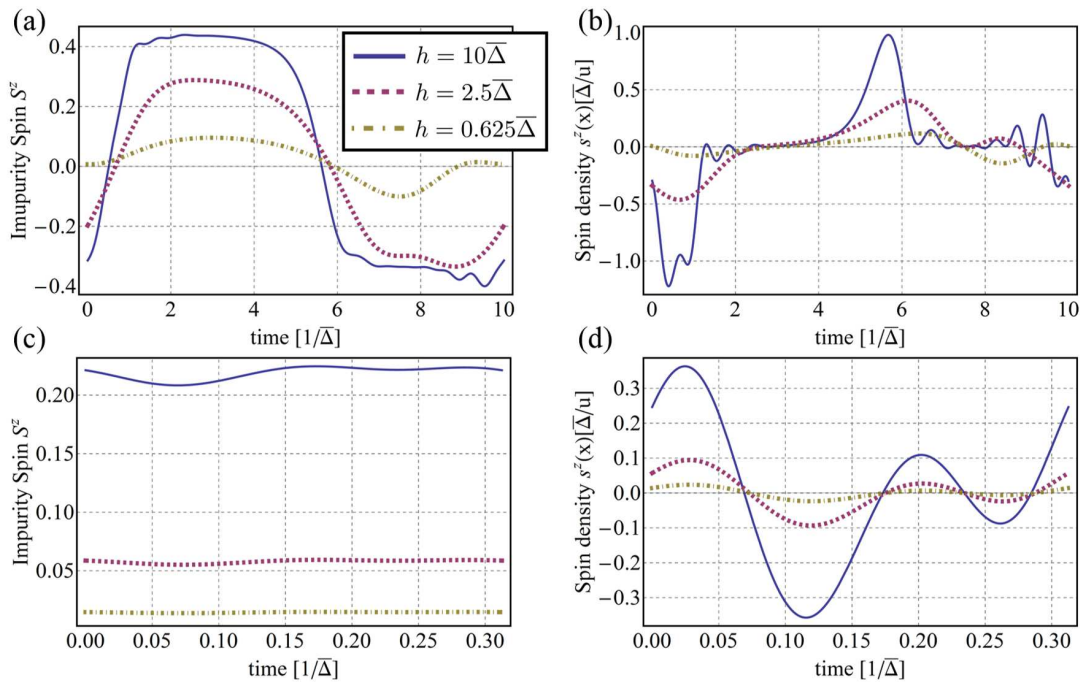}
	\caption{Time evolution of the impurity spin polarization (a), (c) and the spin density of the fermionic bath (b), (d) in a half of the driving period for several choices of local magnetic field $h$ 
			($h=10\overline{\Delta}$, blue solid line; $h=2.5\overline{\Delta}$, red dashed line; $h=0.625\overline{\Delta}$, yellow dashed dotted line). 
			(a) and (b) are the figures for $ \tau = 20/\overline{\Delta} $, while (c) and (d) are for $ \tau = 0.625/\overline{\Delta} $.\label{Fig_InteractionSinusoidalDynamics}}
\end{figure*}

In Fig. \ref{Fig_InteractionSinusoidalDynamics}, we show the time evolution of the impurity spin polarization $ \braket{S^z (t)} $ and the spin density of a fermionic bath $ \braket{s^z(x,t)} \ (x=Mu\tau ,\ M\in \mathbb{N})$ in a half of the driving period. 
Let us define the effective Kondo temperature by $ T_K ^{\eff} = \pi w \overline{\Delta} $ (see Appendix \ref{appendix_b} for the meaning of this definition). 
When the driving period is much longer than the time scale of the inverse of the effective Kondo temperature (see the figures of $ \tau = 20 / \overline{\Delta} $), the system follows a temporal change in the external driving in the time scale of $1/T_K ^{\eff}$. Therefore, the behavior of time evolution is similar to  the case without in-plane interaction driving (see Sec. \ref{Sec_RectangularCase}).
When the driving period is much smaller than the inverse of the effective Kondo temperature (see the figures of $ \tau = 0.625 / \overline{\Delta} $), these values also approach temporally constant values as we have seen for the case without in-plane interaction driving.

It should be noted, however, that the time average of the impurity spin polarization does not vanish even though the local magnetic field does not have the static component (i.e. zeroth mode of Fourier coefficients).
This is because the square of the in-plane interaction parameter becomes small while the local magnetic field is negative.
Namely, the impurity spin is effectively decoupled from the fermionic bath when the local magnetic field is negative.
In the high frequency limit $(\overline{\Delta}/\Omega \rightarrow 0)$, the time average of impurity spin polarization is $\overline{S^z} = \sum _{n \in \mathbb{N}} J_n(h/2\Omega)J_{n+1}(h/2\Omega)$.
This behavior is consistent with the results by Heyl and Kehrein \cite{Heyl2010} in the linear response regime when the coupling term is periodically switched on and off.


\subsection{High frequency limit: analysis of asymptotic steady states}

We now demonstrate that intriguing steady states appear in the high frequency limit. The following analysis is valid for any type of time dependence of $h(t)$ and $J_{\perp}(t)$. Note that the time evolution of the system is determined solely by the function $T_{dk}(t)$ in the steady state (see Eq. (\ref{Eq_TimeEvolution})), so it is sufficient to study the properties of $T_{dk}(t)$.
In the language of the resonant level model, $T_{dk}(t)$ denotes a transition matrix from a state with wave-number $k$ in the fermionic bath to the resonant level. For arbitrary $h(t)$ and $J_{\perp} (t)$, each element of the transition matrix $T_{dk}(t)$ is represented as,
\begin{equation}
	\begin{aligned}
		&T_{dk}(t) = F(t) \sum _{n \in \mathbb{Z}} T_{dk} ^{(n)} (t) \\
		&T_{dk} ^{(n)} (t) = \frac{\tilde{J}^{(n)}}{\sqrt{2\pi \alpha L}} \frac{e^{-i(\epsilon _k -n\Omega)t}}{\epsilon _k +\overline{h}-n\Omega +i\overline{\Delta}},
	\end{aligned}
\end{equation}
where $\overline{f}=\int _0 ^{\tau} f(t) dt /\tau$ is a time average in a period and the coefficients $\tilde{J}^{(n)}$ and $F(t)$ are determined from the Fourier coefficients of the external driving $h(t)$ and $J_{\perp}(t)$.
The transition matrix without driving $T_{dk} ^{\text{bare}}(t)$ in the long time limit is 
\begin{equation}
	T_{dk} ^{\text{bare}} (t) \xrightarrow{t \rightarrow \infty} \frac{J_{\perp}}{\sqrt{2\pi \alpha L}} \frac{e^{-i\epsilon _k t}}{\epsilon _k +h+i\Delta}.
\end{equation}
Thus, $T_{dk} ^{(n)} (t)$ can be interpreted as a "$n$th photon assisted transition matrix".

By taking the high frequency limit $(\overline{\Delta}/\Omega, \ \overline{h}/\Omega \rightarrow 0)$, the impurity-spin polarization $\braket{S^z (t)}$ and the impurity-spin time correlation function, $\braket{S^z (t)S^z(t^{\prime})}$, recover time translational invariance.
We thus end up with the analysic formuae in this limit,
\begin{widetext}
	\begin{equation}
		\begin{aligned}
			&\braket{S_z (t)} \xrightarrow{t \rightarrow \infty, \ \overline{\Delta} /\Omega,\ \overline{h}/\Omega \rightarrow 0} \frac{1}{\pi} \Bigg[ \frac{\Delta ^{(0)}}{\overline{\Delta}} \int _{-\infty} ^{0} + \sum _{n=-1} ^{-\infty} \frac{\Delta ^{(n)}}{\overline{\Delta}} \int _{-\infty} ^{\infty} \Bigg] \frac{d \omega}{(\omega +\overline{h}/{\overline{\Delta}})^2 +1} -\frac{1}{2} \\
			&\begin{aligned}
				&\braket{S_z (t) S_z (t^{\prime})} - \braket{S_z (t)} \braket{S_z (t^{\prime})} \\
				& \begin{aligned}
						\xrightarrow{t,\ t^{\prime} \rightarrow \infty, \ \overline{\Delta} /\Omega,\ \overline{h}/\Omega \rightarrow 0} \Bigg( \Bigg[ \frac{\Delta ^{(0)}}{\overline{\Delta}}& \int _{-\infty} ^{0} +\sum _{n=-1} ^{-\infty} \frac{\Delta ^{(n)}}{\overline{\Delta}} \int _{-\infty} ^{\infty} \Bigg] \frac{d \omega}{\pi} \frac{e^{i\omega \overline{\Delta} (t-t^{\prime})}}{(\omega + \overline{h} / \overline{\Delta})^2+1} \Bigg) \\
																				 & \times \Bigg( \Bigg[ \frac{\Delta ^{(0)}}{\overline{\Delta}} \int _{-\infty} ^{0} +\sum _{n=1} ^{\infty} \frac{\Delta ^{(n)}}{\overline{\Delta}} \int _{-\infty} ^{\infty} \Bigg] \frac{d \omega}{\pi} \frac{e^{i\omega \overline{\Delta} (t-t^{\prime})}}{(\omega - \overline{h} / \overline{\Delta})^2+1} \Bigg) ,
				  \end{aligned}
			 \end{aligned}
		\end{aligned}
		\label{Eq_FastDrivingLimit}		
	\end{equation}
\end{widetext}
where $\Delta ^{(n)} = |\tilde{J}^{(n)}|^2 / 4\pi \alpha u$.
In the equilibrium Kondo model with a local magnetic field $h$ at Toulouse limit and temperature $T$, 
the impurity-spin polarization $ \braket{S^z}_{eq} $ and the impurity-spin time correlation function \cite{Guinea1985,Leggett1978} $ \braket{S^z (t) S^z }_{eq} $ are expressed as follows:
\begin{equation}
	\begin{aligned}
		&\begin{aligned}
			\braket{S_z}_{eq} &= \int _{-\infty} ^{\infty} \frac{d\omega}{\pi} \frac{f(\omega \Delta /T)}{(\omega +h/\Delta)^2 +1} -\frac{1}{2}
		 \end{aligned} \\
		& \begin{aligned}
			&\braket{S_z(t)S_z}_{eq} -\big( \braket{S^z}_{eq} \big) ^2 \\
								& = \Big( \int _{-\infty} ^{\infty} \frac{d\omega}{\pi} \frac{f(\omega \Delta /T)e^{i\omega \Delta t}}{(\omega +h/\Delta)^2 +1} \Big) \Big( \int _{-\infty} ^{\infty} \frac{d\omega}{\pi} \frac{f(\omega \Delta /T)e^{i\omega \Delta t}}{(\omega -h/\Delta)^2 +1} \Big) ,
		\end{aligned}
	\end{aligned}
	\label{Eq_EquilibriumSystem}
\end{equation}
where $f(\epsilon)=1/(e^{\epsilon}+1)$ is the Fermi distribution function.

Comparison between Eq. (\ref{Eq_FastDrivingLimit}) and Eq. (\ref{Eq_EquilibriumSystem}) gives instructive implications for the asymptotic states.
Because $f(\omega \Delta /T) \xrightarrow{T\rightarrow 0} \theta (-\omega)$ and $f(\omega \Delta /T) \xrightarrow{T \rightarrow \infty} \frac{1}{2}$, 
the integrals from $-\infty$ to $0$ in Eq. (\ref{Eq_FastDrivingLimit}) have the properties of the zero temperature state (the time correlation function shows a power law decay $\braket{S^z (t) S^z (t^{\prime})} \propto (t-t^{\prime})^{-2} \ (t\gg t^{\prime})$), while the integrals from $-\infty$ to $\infty$ in Eq. (\ref{Eq_FastDrivingLimit}) have the properties of the infinite temperature state (the time correlation function shows an exponential decay $\braket{S^z(t)S^z(t^{\prime})} \propto e^{-2\overline{\Delta}|t-t^{\prime}|}$).
Thus, the properties of the system are described by {\it a mixture of the zero-temperature state and those of the infinite-temperature state} in general.
This means that the reservoir we consider here cannot prevent the localized spin from heating up completely.
The properties of the zero temperature state come from the bare transitions (i.e. transition processes without photons) while the properties of the infinite temperature come from  the photon-assisted transition processes.
If $\Delta ^{(0)} = 0$ and $\Delta ^{(n)}=\Delta ^{(-n)}$ $(\Delta ^{(n\neq 0)} = 0)$ the properties of the system are described by those at zero (infinite) temperature, respectively. In these special cases, the asymptotic state can be described by the equilibrium state.

Two concrete examples exhibiting either zero- or infinite-temperature properties for asymptotic states are in order here. First, consider the case where the local magnetic field is sinusoidal $h(t)=h\sin (\Omega t)$ and the in-plane interaction is time independent, $\tilde{J}^{(n)} = J_n (h/\Omega)$. Then, the impurity-spin polarization and the impurity-spin time correlation function are exactly reduced to those of the infinite temperature state for the intensities $h$ which satisfy the zeros of the zeroth Bessel function.
This is easily confirmed from Eq. (\ref{Eq_FastDrivingLimit}), where we can put $\Delta^{(0)}=0$. The second example is the $h(t)/\Omega \rightarrow 0$ case: if we take the limit of $h(t)/\Omega \rightarrow 0$, then $\tilde{J}^{(n)} = J_{\perp} ^{(n)}$, where $J_{\perp} ^{(n)}$ is the $n$th Fourier coefficient of $J_{\perp}(t)$. In this case, the steady state cannot be described by the time averaged Hamiltonian even though the frequency of the external driving is set to infinity.
While the impurity-spin polarization and the impurity-spin time correlation function are exactly equivalent to those of the infinite-temperature state when $J_{\perp} ^{(0)} = 0 \ (\Leftrightarrow \overline{J_{\perp}}=0)$,
they are exactly equivalent to those of the zero-temperature state when $J_{\perp} ^{(n)} = 0 \ (\forall n \neq 0) \ (\Leftrightarrow J_{\perp} \text{ is time independent})$.
The other cases cannot be described effectively by any equilibrium systems due to the fluctuation dissipation theorem \cite{Heyl2010}.

To close this section, we make a brief comment on the nonequilibrium Fermi distribution function.
Note that the above asymptotic behaviors originate from the duplication of the Fermi edge (or equivalently duplication of the impurity level) \cite{Tien1963,Kohler2005}.
In calculating observables, we evaluate $\braket{\tilde{c}^{\dagger} _d (t) \tilde{c}_d (t^{\prime})}$, which can be written down explicitly as,
\begin{equation}
	\begin{aligned}
		&\braket{\tilde{c}^{\dagger} _d (t) \tilde{c}_d (t^{\prime})}=\sum _k f(\beta \epsilon _k) \sum_n \Big( T_{dk} ^{(n)} (t) \Big) ^{\ast} T_{dk} ^{(n)} (t^{\prime})\\
		&\propto \sum _k \sum _n |\tilde{J}^{(n)}|^2 f(\beta (\epsilon _k +n\Omega)) \Big( T_{dk} ^{\text{bare}} (t) \Big) ^{\ast} T_{dk} ^{\text{bare}} (t^{\prime})
	\end{aligned}
\end{equation}
The above observation enables us to define the nonequilibrium distribution function $f_{\text{noneq}} (\beta \epsilon) \propto \sum _n |\tilde{J}^{(n)}|^2 f(\beta (\epsilon +n\Omega))$.
The nonequilibrium distribution function has Fermi edges at $\epsilon = n\Omega \ (n\in \mathbb{N})$ and the weight of each edge is $|\tilde{J}^{(n)}|^2$.
Because the Fermi edge at $\epsilon =0$ can pick up the contribution of the density of states of the resonant level and the others cannot, a transition without photons results in the zero-temperature behavior while the others result in the infinite temperature behavior.

\section{Conclusions\label{Sec_Conclusion}}
We have studied the dynamics of a periodically driven anisotropic Kondo model. By carrying out calculations at the Toulouse limit, the exact analytical expressions of the time evolution for an arbitrary periodic time dependence of the local magnetic field and the in-plane interaction have been obtained.

Focusing on the specific case where the local magnetic field is periodically switched on and off, we have specified the time scale of the dynamics and found the intriguing resonance phenomenon between the switching of the local magnetic field and the oscillation of the impurity spin polarization. This resonance results in a characteristic non-monotonic behavior in the time average of the impurity spin polarization as a function of the local magnetic field.

We have also elucidated the intriguing properties emerging in the steady state in the high frequency limit. In that limit, the impurity-spin polarization and the impurity-spin time correlation function recover the time translational invariance. In particular, we have found that the properties of the system are in general described by a mixture of the zero-temperature and infinite-temperate properties, and for special cases, the system can be described either by zero-temperature or infinite-temperature properties. In the latter special cases, the system has the equilibrium-state properties.
If we regard the Kondo model as an open quantum system, the above ``mixture" of states means that the fermionic bath cannot prevent the localized spin from heating up completely. We expect that this behavior may be common to the open quantum systems where the Hamitonian is given by a bilenear form.  Our simple system gives an example which captures the essential structure of such steady states. Extending the present analysis to more generic cases beyond the bilinear Hamitonian is an interesting issue to be explored in the futrure study.

\begin{acknowledgments}
This work was partly supported by a Grand-in-Aid for Scientific Research on Innovative Areas (JSPS KAKENHI Grant No. 15H05855) and also JSPS KAKENHI (No. 25400366).
\end{acknowledgments}

\appendix
\begin{widetext}
\section{Calculation of the time evolution}\label{appendix_a}
Here, we obtain the time evolution of annihilation operators for the time dependent resonant level model:
\begin{equation}
	H(t)=\int dx :\tilde{\psi}^{\dagger}(x) \big( -iu\partial _x \big) \tilde{\psi}(x): 
			+\frac{J_{\perp}(t)}{\sqrt{\mathstrut 2\pi \alpha}} \big( \tilde{\psi}^{\dagger}(0) \tilde{c}_d +h.c. \big) -h(t)\big( \tilde{c}^{\dagger} _d \tilde{c}_d -\frac{1}{2} \big)
\end{equation}
We perform the calculation by using "quadrature by parts". The method is described as follows.
First, divide the time periodic functions $h(t),\ J_{\perp}(t)$ into discrete $M$ time steps, and calculate the time evolution of the system. 
Then, by taking the continuum limit $M\rightarrow \infty$, we obtain the exact expression of time evolution which is applicable for arbitrary $ h(t),\ J_{\perp} (t) $.

Define the discretized functions $h(t),\ J_{\perp}(t)$ as,
\begin{equation}
	\begin{aligned}
		&h(t) = \sum _{N=0} ^{\infty} \sum _{n=1} ^M \theta \big( t-(N-(n-1)/M) \tau \big) \theta \big( (N+n/M)\tau -t \big) h^{(n)} \\
		&J_{\perp}(t) = \sum _{N=0} ^{\infty} \sum _{n=1} ^M \theta \big( t-(N-(n-1)/M) \tau \big) \theta \big( (N+n/M)\tau -t \big) J_{\perp} ^{(n)} \\
		&H^{(n)}=\sum _k uk :\tilde{c}^{\dagger} _k \tilde{c}_k : + \frac{J_{\perp} ^{(n)}}{\sqrt{2\pi \alpha L}} \sum _k \big( \tilde{c}^{\dagger} _k \tilde{c}_d +h.c. \big) -h^{(n)} \big( \tilde{c}^{\dagger} _d \tilde{c}_d -\frac{1}{2} \big)
	\end{aligned}
\end{equation}
The time evolution by the Hamiltonian $H^{(n)}$ is given by
\begin{equation}
	\tilde{c}_l (t) = e^{iH^{(n)}t}\tilde{c}_l e^{-iH^{(n)}t} = \sum _{l^{\prime}} G^{(n)} _{ll^{\prime}} (t) \tilde{c}_{l^{\prime}},\ (l,\ l^{\prime} = k,\ d),
\end{equation}
where the transition matrices $G^{(n)} _{ll^{\prime}} (t)$ are
\begin{equation}
	\begin{aligned}
		&G^{(n)} _{dd} (t) = e^{ih^{(n)}t-\Delta ^{(n)} t},\ \Delta ^{(n)} = \frac{\big(J_{\perp} ^{(n)}\big) ^2}{4\pi \alpha u} ,\ V^{(n)} =\frac{J_{\perp} ^{(n)}}{\sqrt{\mathstrut 2\pi \alpha L}} \\
		&G^{(n)}_{kd}(t)=G^{(n)}_{dk}(t)=V^{(n)} \frac{e^{-i\epsilon _k t}-G_{dd} ^{(n)} (t)}{\epsilon _k +h^{(n)}+i\Delta ^{(n)}} \\
		&G^{(n)}_{kk^{\prime}} (t) = \delta _{k,k^{\prime}} e^{-i\epsilon _k t} + Q_{kk^{\prime}} ^{(n)} (t) \\
		&Q^{(n)}_{kk^{\prime}} (t) = \Big( V^{(n)} \Big) ^2 \Bigg[ \frac{1}{\epsilon _k -\epsilon _{k^{\prime}}} \Big( \frac{e^{-i \epsilon _k t}}{\epsilon _k +h^{(n)} +i\Delta ^{(n)}} -\frac{e^{-i \epsilon _{k^{\prime}}t}}{\epsilon_{k^{\prime}} +h^{(n)} +i\Delta ^{(n)}} \Big) 
										+ \frac{G^{(n)} _{dd} (t)}{(\epsilon_k +h^{(n)} +i\Delta ^{(n)})(\epsilon_{k^{\prime}} +h^{(n)} +i\Delta ^{(n)})} \Bigg].
	\end{aligned}
\end{equation}
The time evolution in a period is obtained as,
\begin{equation}
	\begin{aligned}
		\tilde{c}_l (\tau) &= e^{i H^{(1)} \tau /M} \cdots e^{i H^{(M)} \tau /M} \tilde{c}_l e^{-i H^{(M)} \tau /M} \cdots e^{-i H^{(1)} \tau /M}\\
						   &= \sum _{\{ l_i \} }G^{(M)} _{ll_1} \big(\tau /M \big) G^{(M-1)} _{l_1 l_2} \big(\tau /M \big) \cdots G^{(1)} _{l_{M-1} l_M} \big(\tau /M \big) \tilde{c}_{l_M}\\
						   &\equiv \sum _{l^{\prime}} M_{ll^{\prime}} \tilde{c}_{l^{\prime}}.
	\end{aligned}
\end{equation}
Thus, simultaneous ordinary differential equations with time dependent coefficients can be mapped onto a problem of multiplication of matrices.
This multiplication of matrices is easily done in this system due to a linear dispersion.
For example, in the calculation of
\begin{equation}
	\sum _l G^{(n)} _{dl} G^{(n+1)} _{ld} = \sum _k G^{(n)} _{dk} G^{(n)} _{kd} + G^{(n)} _{dd} G^{(n+1)} _{dd},
\end{equation}
the first term disappears because the function $G^{(n)} _{dk}$ does not have any poles. 
The other calculations can be performed in the same way.
The transition matrices $T_{ll^{\prime}} (t) $ at each time step $t=n \delta \tau $ are
$T_{ll^{\prime}} (n \delta \tau) = \sum _{ \{ l_i \} } G^{(n)} _{ll_1} (\delta \tau) G^{(n-1)} _{l_1 l_2} (\delta \tau) \cdots G^{(1)} _{l_{n-1} l^{\prime}} (\delta \tau)$, and thus we end up with the explicit expressions for $n>1$,
\begin{equation}
	\begin{aligned}
		& T_{dd} (n\delta \tau) = \prod _{m=1} ^{n} G^{(m)} _{dd} (\delta \tau) \\
		& T_{kd} (n\delta \tau) = \sum _{m=0} ^{n-1} e^{-im\epsilon _k \delta \tau} G^{(n-m)} _{kd} (\delta \tau) T_{dd} ((n-m-1)\delta \tau)\\
		& T_{dk} (n\delta \tau) = \sum _{m=1} ^{n-1} \Big( \prod _{l=n-m+1} ^n G^{(l)} _{dd} (\delta \tau) \Big) G^{(n-m)} _{dk} (\delta \tau) e^{-i(n-m-1)\epsilon _k \delta \tau} + G^{(n)} _{dk} (\delta \tau) e^{-i(n-1)\epsilon _k \delta \tau} \\
		& T_{kk^{\prime}} (n\delta \tau) = \delta _{k,k^{\prime}} e^{-in \epsilon _k \delta \tau} + P_{kk^{\prime}} (n \delta \tau) \\
	\end{aligned}
\end{equation}
Then, by taking the continuum limit and carrying out some calculations, we obtain the time evolution of operators in the steady state:
\begin{equation}
	\begin{aligned}
		& \tilde{c}_d (t) = \sum _k T_{dk} (t) \tilde{c}_k ,\ \tilde{\psi}(x,t)=\tilde{\psi}(x-ut)-i\sum _k \frac{J_{\perp}(t-x/u)}{\sqrt{2\pi \alpha u^2}} \theta(x) T_{dk} (t-x/u) \tilde{c}_k \\
		& T_{dk} (t) = \frac{-ie^{-iukt}}{1-e^{iuk\tau}M_{dd}} \int _0 ^{\tau} ds \frac{J_{\perp}(t-s)}{\sqrt{2\pi \alpha L}} \exp \Big[ i\int _{\tau -s} ^{\tau} \big( h(s^{\prime}+t) -\Delta (s^{\prime}+t) \big)ds^{\prime} +iuks \Big] \\ 
		& M_{dd} = \exp \Big[ \int _0 ^{\tau} \big( ih(t) -\Delta (t) \big) dt \Big] ,\ \Delta (t)= \frac{J_{\perp} ^2 (t)}{4\pi \alpha u}.
	\end{aligned}\label{Eq_TimeEvolution}
\end{equation}


\section{Effective Kondo temperature}\label{appendix_b}

The effective Kondo temperature in the steady state when the the local magnetic field is absent ($ h(t)=0 $) can be defined by the impurity spin susceptibility \cite{Heyl2010}.
At zero temperature in equilibrium, the imaginary part of the Fourier coefficient of the impurity-spin susceptibility $ \chi _{eq} (t) = i \theta (t) \braket{ [ S^z (t), S^z ] } _{eq} $ has a characteristic peak structure near $ \omega \sim T_K $ reflecting the existence of Kondo singlet.
The imaginary part of the Fourier coefficient of the impurity spin susceptibility in equilibrium is given by \cite{Guinea1985,Leggett1978}
\begin{equation}
	\Im \tilde{\chi}^{T=0} _{eq} (\omega, T_K) = \frac{1}{2\pi} \frac{1}{1+\big( \pi w \omega / T_K \big) ^{2}/4} \Big[ \frac{1}{\omega} \log \Big( 1+\Big( \frac{\pi w \omega}{T_K} \Big) ^2 \Big)
																																		 +\frac{\pi w}{T_K} \Arctan \Big( \frac{\pi w \omega}{T_K} \Big) \Big] .
\end{equation}
Substituting $ h(t) = 0 $ and $\tilde{J}^{(n)}=J_{\perp} ^{(n)}$ to Eq. (\ref{Eq_FastDrivingLimit}), the following relation is obtained:
\begin{equation}
	\Im \tilde{\chi} (\omega) = \frac{\Delta ^{(0)}}{\overline{\Delta}} \Im \tilde{\chi}^{T=0} _{eq} (\omega,\pi w \overline{\Delta})
\end{equation}
Thus, in this steady state, the effective Kondo temperature defined as $T_K ^{\text{eff}} := \pi w \overline{\Delta}$ is determined by the time average of the square of the in-plane interaction parameter (in equilibrium state, $T_K = \pi w \Delta$).
Because the prefactor is different by a factor $ \Delta ^{(0)} /\overline{\Delta} $, the response to the local magnetic field is supressed by $ \Delta ^{(0)} /\overline{\Delta} $.

\end{widetext}

\bibliography{manuscript}

\end{document}